\documentclass[showpacs,prl,aps,superscriptaddress,twocolumn,floatfix]{revtex4}
\usepackage{graphicx,float,amsmath,amssymb}

\begin{document}

\title{Superfluid Motion of Light}

\author{Patricio Leboeuf}
\author{Simon Moulieras}
\affiliation{Laboratoire de Physique Th\'eorique et Mod\`eles Statistiques, CNRS, 
Universit\'e Paris Sud, UMR8626, 91405 Orsay Cedex, France}
\begin{abstract}
Superfluidity, the ability of a fluid to move without dissipation, is one of the most spectacular manifestations of the quantum nature of matter. We explore here the possibility of superfluid motion of light. Controlling the speed of a light packet with respect to a defect, we demonstrate the presence of superfluidity and, above a critical velocity, its breakdown through the onset of a dissipative phase. We describe a possible experimental realization based on the transverse motion through an array of waveguides. These results open new perspectives in transport optimization.
\end{abstract}


\maketitle


Next year will bring the opportunity to celebrate the 100th anniversary of the discovery of superconductivity \cite{onnes}. This remarkable property is often related to a more fundamental phenomenon, the Bose Einstein condensation, where a single quantum state is occupied by a macroscopic number of particles. The bosons that condense may be coupled electrons that form Cooper pairs, as in superconducting metals \cite{meissner}, atoms \cite{cornell}, like in the original experiments in superfluid (SF) $ ^{4} He$ \cite{kap}, or molecules \cite{zwi}. They may also be formed of more complex particles, like fermionic atom pairs \cite{regal}, or polaritons, a composite of a photon and an exciton \cite{kasp}. Superfluidity of polaritons in semiconductor cavities was explicitly tested recently \cite{amo}.

There are different definitions of superfluidity, each one may emphasize a particuler physical aspect. Here superfluidity means the existence of a finite critical velocity $v_c >0$ below which the motion of the fluid is dissipationless. A particularly simple way to experimentally implement this test is by moving through the fluid an obstacle, or localized external potential. When the potential is weak, it has been shown long ago \cite{landau} that $v_c = c_s$, where $c_s$ is the speed of sound in the fluid \cite{foot1}. Above the critical velocity, superfluidity is broken and dissipative effects appear. 

It is important to note that a finite critical velocity is directly related to the presence of interactions between the bosons. The interactions control the long wavelength structure of the dispersion relation of the fluid. In a mean field approximation, weakly interacting bosons may be modeled, with good accuracy, by the Gross-Pitaevsky equation. The latter corresponds to a Schr{\"o}dinger equation with an additional nonlinear term that describes the interactions. In particular, this equation reproduces correctly the Bogoliubov dispersion relation mentioned above for the excitations of the interacting fluid. Interestingly, when the propagation of light is considered through a nonlinear medium of the Kerr-type uniform in one direction, in the paraxial approximation a similar equation is obtained for the slowly varying envelope of the optical field of a given wavenumber and frequency. The analogy between the Gross-Pitaevsky equation and the light propagation in nonlinear media has been exploited in the past to test basic quantum effects with optics, like Bloch oscillations \cite{pertsch,christo} or Anderson localization \cite{schwartz}. Because of the similarities of the two equations, and since the Gross-Pitaevsky equation predicts SF motion, it is natural to push further the analogies and consider the possibility to observe a new state of light, e.g. superfluidity in an optical nonlinear medium. Based on a self-defocusing refractive medium inside a Fabry-P\'erot cavity, an optical analog of a SF has indeed been proposed \cite{chiao}. However, the results of the numerical simulations based on a transient regime were not conclusive. Moreover, no clear evidence of the existence of a SF critical velocity was provided. Therefore, the existence of photonic superfluidity in nonlinear media, as well as its experimental observation, remain open issues.

Our purpose here is (i) to provide clear evidence of SF motion of light in a nonlinear medium as well as of its breakdown and (ii) propose an experiment that allows the observation of these effects. For simplicity, we will focus on the propagation of light in an effective one-dimensional array of waveguides.

\begin{figure}
\includegraphics*[width=0.85\linewidth,angle=0]{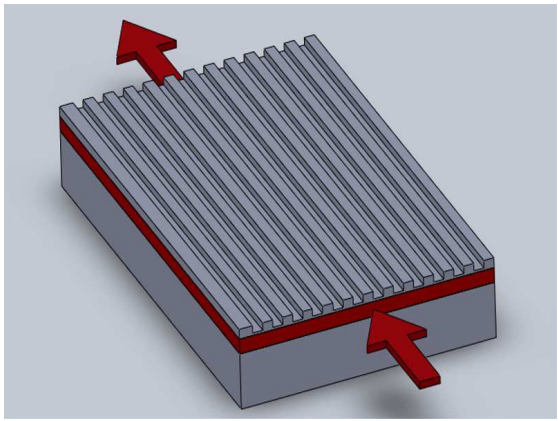}
\caption{(color online) Schematic view of an array of waveguides. The red arrows indicate the input beam and the output facet, where the light intensity distribution is measured.} \label{F1}
\end{figure}

In these materials, light propagates in a medium where the refractive index has been spatially modulated. A typical set up consists of a periodic modulation of a two-dimensional layer, where an array of equally spaced identical waveguides is formed (see Fig.\ref{F1}). Outside each of the waveguides the optical field intensity decreases exponentially. When the distance is such that the overlap between the fields of neighbouring waveguides is small, the optical tunnelling between adjacent guides may simply be modelled by a hopping term. Light propagates along the guides in the longitudinal direction and hops from guide to guide in the transverse direction. Moreover, the width of each waveguide may be engineered in order to modify the energy of the local (quasi) bound state light mode, and Kerr materials may be used to include nonlinear effects. Under these conditions, the optical field amplitude $A_k$  of light at the $k^{th}$ lattice site (or waveguide) obeys the following discrete nonlinear Schr\"odinger equation \cite{christo} (in the paraxial approximation):
\begin{equation}
i\frac{\partial A_k}{\partial z} = -C \left( A_{k+1} + A_{k-1} \right) + \gamma |A_{k}|^2 A_{k} + \epsilon_k A_{k}
\label{DNLS}
\end{equation}
where $C$ is the tunneling rate between two adjacent sites, $\epsilon_k$  the on-site energy, and $\gamma >0$ the strength of the self-defocusing nonlinearity of the medium. The left hand side describes the propagation along the longitudinal $z$-axis of the waveguide, and replaces time in the Schr\"odinger equation. We measure all distances in units of the incident light wavenumber, hence $z$, $C$, $\epsilon_k$, and $\gamma |A_{k}|^2$ are dimensionless.

The possibility to engineer the different characteristics of the array make these photonic lattices unique in their ability to control the flow of light. A laser beam is shone on the input facet of an array of $N$ waveguides, propagates across the photonic lattice, to finally reach the output facet where the intensity distribution is measured (as shown in Fig.\ref{F1}). The length of the structure in the z-direction thus determines the time of propagation across the lattice.

In order to test superfluidity of light, we are interested in the scattering properties of an incident pulse on a localized defect. In the absence of a defect, the light pulse spreads in a way that strongly depends on the nonlinear coefficient. We are not interested in this free propagation, which was studied in detail in the past \cite{christo}. Ideally, we would like to analyze the propagation of a packet whose shape is, in the absence of the obstacle, independent of time, in order to clearly single out the influence of the defect on its propagation. There are different ways to realize this. One possibility, which is easily accessible experimentally, is to control the on-site energies $\epsilon_k$ by modulating the width of each waveguide to build a harmonic confining potential, $\epsilon_k=\epsilon_0 + \frac{1}{2}\omega^2 k^2$, where $\epsilon_0$ is the reference on-site energy, and $\omega$ measures the frequency in units of normalized $1/z$ \cite{foot4}. The site $k=0$ defines the center of the lattice. The advantages of such a set up are multiple. One can shine on the lattice a light packet whose center is located at an arbitrary distance $d$ from the bottom of the potential. As it propagates in the longitudinal direction, the packet will oscillate in the transverse direction with frequency $\omega$ \cite{foot2}. One can show that for $\gamma >0$ the frequency $\omega$ coincides with that of $\gamma =0$ \cite{kohn}. Moreover, the shape of the packet does not vary in time if initially it is given by the (translation) of the stationary ground state solution of Eq.(\ref{DNLS}). Thus, with a positive nonlinearity, such a light packet oscillates with a frozen shape, and its velocity at the bottom of the potential is $v= \omega \cdot d$. This allows to control the transverse speed of light.

We now include the defect at the center of the harmonic potential. A simple way to experimentally implement it is by a local variation of the on-site energy, $\epsilon_k=\epsilon_0 + \frac{1}{2}\omega^2 k^2 + U_0 \delta_{k,0}$ (where $U_0$ represents the defect strength). The purpose now is to study, for different relative velocities $v$, the oscillations of the light packet in the presence of the defect. If the light is scattered by the defect, dissipative processes are induced that transform the coherent collective oscillation into disordered fluctuations of the light intensity. As a consequence a damping of the collective character of the oscillations is expected. On the contrary, if SF motion occurs, the light pulse is able to move through the defect without losing collectivity (e.g., without changing its global shape). It only creates a local intensity depletion around the defect \cite{PRA2001}. By analogy with the dispersion relation of the corresponding (continuous) Gross-Pitaevsky equation, Eq.(\ref{DNLS}) predicts a SF motion for velocities below a critical threshold $v_c$ which, for weak perturbations, is of the order of $\sqrt{C \gamma |A|^2}$, where $|A|^2$ is the light intensity at the center of the light pulse. For typical experimental set ups \cite{eis}, this velocity is $v_c \approx 2 \cdot 10^{-2}$, which corresponds, in the original units, to the ratio of the transverse speed to the speed of light in the photonic structure.

\begin{figure}
\includegraphics*[width=1.0\linewidth,angle=0]{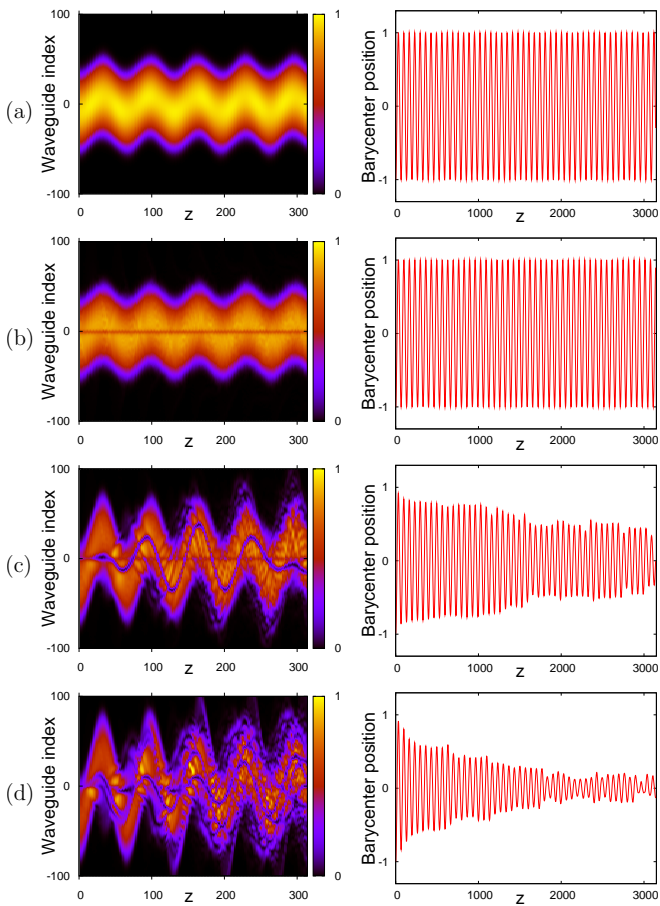}
\caption{(color online) Time evolution of a light pulse propagating through an array of waveguides modulated by a harmonic on-site energy, with parameters $C=100$, $\omega=0.1$, $\gamma=80$, $N=300$. Left column: intensity, normalized to the maximal one. Right column: amplitude of the oscillatory motion of the light packet barycenter
(for the first $ \approx 50$ oscillations; the amplitude is normalized to the initial value). Top panel (a): free motion. Following panels (b)-(d): oscillation in presence of an additional obstacle located at the center of the lattice. (b): SF motion. (c): speed slightly above the critical velocity. (d): speed well above the critical velocity.}
\label{F2}
\end{figure}

Figure \ref{F2} shows, for different initial positions $d$, the first oscillations of the light packet. In the absence of an obstacle (Fig.\ref{F2}a), the packet oscillates freely with constant shape and amplitude. In presence of the defect and for small amplitudes (Fig.\ref{F2}b), no damping or dissipative process is observed (see right column of the Figure). The only manifestation of the presence of the defect on the light pulse is a local intensity depletion at the position of the defect, which is clearly visible in Fig.\ref{F2}b as a horizontal red line. The motion is qualitatively similar to the free oscillation of the light pulse, shown in Fig.\ref{F2}a, aside from a slight modification of the frequency, that can be explained theoretically \cite{mathias}. Increasing the amplitude and thus the relative velocity with respect to the defect, there is a critical speed above which the shape of the light pulse is qualitatively modified as it propagates (Fig.\ref{F2}c). What is observed is, in particular, the emission of grey soliton-like perturbations (dark blue trajectory), which detach from the defect and travel across the light packet with a non trivial dynamics. This dissipative process produces a damping of the oscillations. As the velocity is further increased (Fig.\ref{F2}d), the shape of the light pulse is subjected to massive deformations through phonon-like and solitonic emissions. The complex dynamics of the excitations signals the onset of a strong dissipative process that destroys the collectivity of the oscillations, thus leading to a strong damping.

\begin{figure}
  \includegraphics[width=1.0\linewidth]{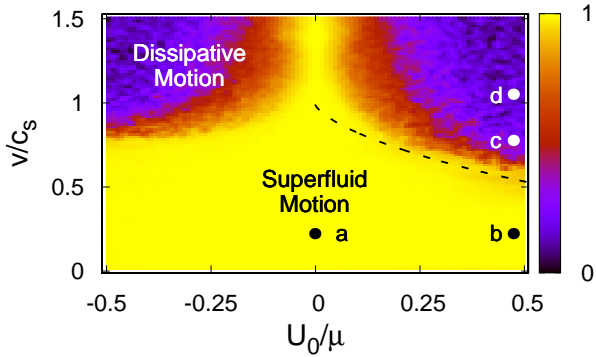}
  \caption{(color online) Fluidity factor computed after a time $z_f= 50 \cdot 2\pi /\omega$, as a function of the normalized strength of the defect and transverse speed of the light pulse. Yellow regions correspond to SF motion, whereas damping is observed in the darker regions. The dashed line represents the analytical non-perturbative prediction of the critical velocity (see text). The full dots $(a)-(d)$ indicate the parameters used in Fig.\ref{F2} ($(a)$: $U_0=0$, $v/c_s=0.23$; $(b)$, $(c)$ and $(d)$ : $U_0/\mu=0.48$ and $v/c_s=0.23$, $0.76$ and $1.05$, respectively).}
  \label{diag}
\end{figure}

A quantitative way to characterize the dissipative process observed in Fig.\ref{F2} is to numerically evaluate the fluidity factor, defined as the ratio of the amplitude of the oscillation around some final time, to the initial amplitude, $\Sigma_k k |A_k|^2/\Sigma_k k |A_{k}^0|^2$, where the $A_k^{0}$'s are the incident amplitudes of the field. 
This factor varies from $0$ for a totally damped motion to $1$ for an undamped one.
We show in Fig.\ref{diag} the computation of the fluidity factor for different velocities and different attractive ($U_0<0$) and repulsive ($U_0>0$) defect strengths. The final time corresponds to $50$ free oscillation periods. The velocities are normalized to the perturbative critical velocity, defined as $c_s=\sqrt{2 C \mu}$, where $\mu$ is the chemical potential of the incident light packet, $\mu= \Sigma_k {A_k^{0}}^{*} \{-C \left( A_{k+1}^0 + A_{k-1}^0 \right) + \gamma |A_{k}^0|^2 A_{k}^0 + \frac{1}{2} (\omega k)^2 A_{k}^0 \}$. At low velocities, the fluidity factor is equal to one and the light pulse presents a perfect transmission through the scattering potential. A local peak (resp. dip) is observed on the light intensity when the defect is attractive (resp. repulsive), but this does not affect qualitatively the dynamics of the oscillations. This demonstrates that the transverse motion of the light is superfluid for a well defined parameter range in transverse speed of light and defect strength. As the velocity increases a sharp transition towards a phase of damped dynamics is observed. This border defines the critical velocity $v_c$. Above $v_c$, non local dissipative excitations are allowed, and superfluidity breaks down. As shown in Fig.\ref{diag}, $v_c$ coincides with the perturbative one, $c_s$, for weak defect strengths. However, as the strength increases, $v_c$ deviates from the perturbative estimate. A non-perturbative analysis is required to describe the threshold \cite{PRA2001,mathias}. That analysis also allows to explained the dissymmetry observed in Fig.\ref{diag} between positive and negative values of $U_0$. The critical velocity as a function of the strength of the defect, shown in Fig.\ref{diag}, is computed by solving the equation $U_0/\mu=K(v_c(U_0)/c_s)$ where $K(x)=\frac{1}{2\sqrt{2}x} (-8x^4-20 x^2 + 1 +(1+8x^2)^{3/2})^{1/2}$.

The method used here to test superfluidity of light has a close counterpart in the physics of ultracold atoms.
The influence of a local potential on the flow of a condensate \cite{ono,PRA2001} or on the damping of dipole oscillations \cite{fort,mathias,hulet} became in recent years an important and accepted experimental and theoretical tool to analyze the dynamics of ultracold Bose-Einstein condensates, in particular as a test of superfluidity. 

We have considered here nonlinear optics at a purely classical level. In the paraxial approximation, its formal description is identical to the Gross-Pitaevsky equation, which describes a mean-field dilute condensate of atoms. Many open issues, related to the underlying microscopic quantum theory of light and to its connections with the phenomenon of Bose-Einstein condensation, deserve further investigation \cite{chiao}.
In a wider context, the physics described here is similar to that observed in the wave resistance of a moving disturbance at the surface of a liquid \cite{PGDG}, or to the Cherenkov radiation of a charged particle moving through a dielectric medium \cite{cherenkov}.

To conclude, we have shown that, in a typical experiment, for transverse speeds of the order of $10^{-2}$ the speed of light in a self-defocusing nonlinear medium the light motion becomes superfluid. In contrast, superfluidity does not occur in a focusing ($\gamma<0$) medium. The effect described is not inherent to discrete lattice structures, and is expected to occur in continuous media as well, provided the refraction index can be carefully designed. The main interest of the set up based on an array of waveguides is the ability to easily control the different parameters.  Furthermore, in our effective one-dimensional geometry we have also identified the emission of solitonic and phonon-like excitations as the main mechanisms that contribute to the breakdown of the SF motion above the critical velocity. In two dimensions superfluidity is also expected to occur, with a transition to a dissipative flow related to the emission of optical vortices. We believe the SF motion described here is a general property, which may be observed for an arbitrary scattering potential (not limited to a localized defect). The propagation in the presence of, e.g., random fluctuations of the on-site energies or of the refraction index is of particular interest, since randomness is inherent to any fabrication process. In analogy with similar recent studies in the physics of ultracold atoms \cite{tobias,hulet}, one may expect the existence of SF motion of light in presence of disorder.

Acknowledgements: We thank M. Albert, N. Pavloff, and Y. Lahini for fruitful discussions and T. Paul for providing us a nonlinear Schr\"odinger equation program.


\begin{thebibliography}{99}
\bibitem{onnes} H. K. Onnes, Commun. Phys. Lab. Univ. Leiden {\bf 12}, 120 (1911).
\bibitem{meissner} W. Meissner and R. Ochsenfeld, Naturwiss. {\bf 21} 787 (1933).
\bibitem{cornell} M. H. Anderson {\it et al.}, Science, New Series, {\bf 269}, 198 (1995).
\bibitem{kap} P. Kapitza, Nature {\bf 141}, 74 (1938). J. F. Allen and A. D. Misener, Nature {\bf 141}, 75 (1938).
\bibitem{zwi} M. W. Zwierlein {\it et al.}, Phys. Rev. Lett. {\bf 91}, 250401 (2003).
\bibitem{regal} C. A. Regal, M. Greiner, and D. S. Jin, Phys. Rev. Lett. {\bf 92}, 040403 (2004).
\bibitem{kasp} J. Kasprzak {\it et al.}, Nature {\bf 443}, 409 (2006).
\bibitem{amo} A. Amo {\it et al.}, Nature Physics {\bf 5}, 805 (2009).
\bibitem{landau} L. Landau , Phys. Rev. {\bf 60}, 356 (1941).
\bibitem{foot1} Roton modes are not relevant for our purpose and will henceforth be ignored.
\bibitem{pertsch} T. Pertsch, P. Dannberg, W. Elflein, and A. Bräuer, Phys. Rev. Lett. {\bf 83}, 4752 (1999). 
\bibitem{christo} D. N. Christodoulides, F. Lederer and Y. Silberberg, Nature {\bf 424}, 817 (2003).
\bibitem{schwartz} T. Schwartz, G. Bartal, S. Fishman, and M. Segev, Nature {\bf 446}, 52 (2006); Y. Lahini {\it et al.}, Phys. Rev. Lett. {\bf 100}, 013906 (2008).
\bibitem{chiao} R. Y. Chiao, and J. Boyce, Phys. Rev. A {\bf 60}, 4114 (1999); R. Y. Chiao, Opt. Commun. {\bf 179}, 157 (2000); E. L. Bolda, R. Y. Chiao, and W. H. Zurek, Phys. Rev. Lett. {\bf 86}, 416 (2001).
\bibitem{foot4} another possibility to create such a potential is to consider the motion through uniformly twisted arrays of waveguides, see S. Longhi, Phys. Rev. B {\bf 76}, 195119 (2007).
\bibitem{foot2} Although quite interesting, the amplitude of the oscillations used here are sufficiently small in order to ignore the impact on the dynamics due to the discrete aspects of the lattice.
\bibitem{kohn} W. Kohn, Phys. Rev. {\bf 123}, 1242 (1961); L. Vichi and S. Stringari, Phys. Rev. A {\bf 60}, 4734 (1999).
\bibitem{PRA2001} P. Leboeuf and N. Pavloff, Phys. Rev. A {\bf 64}, 033602 (2001).
\bibitem{eis} H. S. Eisenberg, Y. Silberberg, R. Morandotti, A. R. Boyd and J. S. Aitchison, Phys. Rev. Lett. {\bf 81}, 3383 (1998).
\bibitem{mathias} M. Albert, T. Paul, N. Pavloff and P. Leboeuf, Phys. Rev. Lett. {\bf 100}, 250405 (2008).
\bibitem{ono} R. Onofrio {\it et al.}, Phys. Rev. Lett. {\bf 85}, 2228 (2000); A. P. Chikkatur et al., Phys. Rev. Lett. 85, 483 (2000); S. Inouye et al., Phys. Rev. Lett. 87, 080402 (2001); P. Engels and C. Atherton, Phys. Rev. Lett. {\bf 99},  160405 (2007).
\bibitem{fort} C. Fort {\it et al.}, Phys. Rev. Lett. {\bf 95}, 170410 (2005).
\bibitem{hulet} D. Dries, S. E. Pollack, J. M. Hitchcock and R. G. Hulet, arXiv:1004.1891.
\bibitem{PGDG} E. Rapha\"el and P. G. de Gennes, Phys. Rev. E {\bf 53}, 3448 (1996).
\bibitem{cherenkov} P. A. Cherenkov, C. R. Acad. Sci. URSS {\bf 8}, 451 (1934). I. Frank and I. Tamm, C. R. Acad. Sci. URSS {\bf 14}, 109 (1937).
\bibitem{tobias} T. Paul, P. Schlagheck, P. Leboeuf, and N. Pavloff ,Phys. Rev. Lett. {\bf 98}, 210602 (2007).T. Paul, M. Albert, P. Schlagheck, P. Leboeuf and N. Pavloff, Phys. Rev. A {\bf 80}, 033615 (2009).
\end{thebibliography}
\end{document}